\documentclass[10pt,twocolumn,letterpaper]{article}

\usepackage[pagenumbers]{cvpr}
\usepackage{times}
\usepackage{epsfig}
\usepackage{graphicx}
\usepackage{amsmath}
\usepackage{amssymb}

\usepackage{booktabs}
\usepackage{array}
\usepackage{multirow}
\usepackage{xcolor}
\usepackage{url}
\usepackage[hidelinks]{hyperref}
\hypersetup{
  pdftitle={Which Acoustic Features Survive Compression? Tracker-Estimated First-Formant Statistics for Codec-Robust Audio Deepfake Detection},
  pdfauthor={Yash Sukhdeve, Aaron Storey, Masudul H. Imtiaz},
  pdfsubject={Preprint; accepted at IEEE/IAPR IJCB 2026; not published in the conference proceedings},
  pdfkeywords={audio deepfake detection, formants, codec robustness, voice anti-spoofing}
}

\graphicspath{{figs/}}

\begin{document}

\title{Which Acoustic Features Survive Compression?\\
Tracker-Estimated First-Formant Statistics for Codec-Robust Audio Deepfake Detection}

\author{Yash Sukhdeve$^{1,*}$ \quad Aaron Storey$^{2}$ \quad Masudul H. Imtiaz$^{1}$\\[3pt]
{\small $^{1}$Department of Electrical and Computer Engineering}\\
{\small $^{2}$Department of Computer Science}\\
{\small Clarkson University, Potsdam, NY 13699, USA}\\[2pt]
{\tt\small \href{mailto:sukhdeya@clarkson.edu}{sukhdeya@clarkson.edu}\quad
\href{mailto:astorey@clarkson.edu}{astorey@clarkson.edu}\quad
\href{mailto:mimtiaz@clarkson.edu}{mimtiaz@clarkson.edu}}\\[2pt]
{\small $^{*}$Corresponding author: Yash Sukhdeve}
}

\AddToShipoutPicture*{\AtPageLowerLeft{%
  \put(3,113.604){\rotatebox{90}{%
    \fontsize{12bp}{14.4bp}\selectfont\bfseries
    Accepted at \textit{IEEE/IAPR} International Joint Conference on Biometrics (IJCB), 2026%
  }}%
}}

\maketitle
\thispagestyle{plain}


\begin{abstract}\normalfont
Audio deepfake detectors that work well on clean lab recordings often degrade after telephone or platform compression. This paper asks a specific question: which low-dimensional acoustic descriptors stay informative after codec degradation? We screen 36 utterance-level features (formants, PNCCs, LFCCs, prosody) on ASVspoof~2019 data and test on the official ASVspoof~2021 LA set across seven codec conditions. Of 33 statistically significant descriptors, only four show large practical separation ($|r|>0.5$, bootstrap 95\% CIs), all first-formant (F1) measures, highly redundant (PC1: 79\% variance). The payoff is stability: the 36-feature SVM degrades 14\,pp under codec mismatch (9.5\%$\to$23.5\% EER), while the four-feature RBF-SVM degrades only 5.9\,pp (17.0\%$\to$22.9\%). The F1 subset varies 3.4\,pp across conditions versus 13.1\,pp for the full set. F1 differences persist after controlling for speaker identity ($p<10^{-10}$, mixed-effects model). The results suggest that physically interpretable F1 statistics offer a compact, codec-stable diagnostic signal that may complement higher-accuracy neural detectors.
\end{abstract}

\section{Introduction}
Speech synthesis and voice conversion now produce convincing fake speech, creating risks for fraud, impersonation, and fabricated audio evidence. The detection literature has grown fast, from cepstral front ends and spectrogram-based neural networks to self-supervised and style-aware architectures \cite{li2024survey,guo2024wavlm,zhu2024slim}. But one problem recurs: detectors trained on clean audio lose accuracy once the signal passes through phone networks, streaming platforms, or messaging apps that apply lossy compression \cite{yamagishi2021asvspoof,liu2023asvspoof,chen2021ur}.

This mismatch matters most in forensic and biometric settings: a model trained on studio-quality data exploits subtle cues that separate spoofing systems from real recordings, but once the audio passes through a codec those cues are distorted or erased. Yet few studies have measured how much each feature degrades across specific codec conditions, or which features survive at all.

Unlike systems that optimize benchmark EER, we ask which hand-crafted acoustic descriptors keep their discriminative power after codec degradation. The motivation is forensic: when a deepfake is entered as evidence or flagged in a fraud investigation, someone has to explain why the audio is fake. A neural embedding cannot do that. A formant statistic can. Compact, interpretable descriptors also reveal which parts of the speech signal survive deployment, exposing failure modes hidden inside high-dimensional embeddings. Throughout, we treat the four-feature F1 model as an interpretable diagnostic/forensic signal rather than a standalone countermeasure: its value lies in codec stability and explainability, not in competitive detection accuracy.

The study uses clean training data from ASVspoof 2019 LA and the official ASVspoof~2021~LA evaluation set (seven codec/channel conditions), comparing four feature families: formants, power-normalized cepstral coefficients (PNCCs), linear-frequency cepstral coefficients (LFCCs), and prosody. Out of 36 utterance-level descriptors, 33 are statistically significant after Benjamini--Hochberg correction, but only four show large practical effect size ($|r|>0.5$), and all four are first-formant (F1) measures. When we restricted the detector to those four cues, EER varied only 3.4~percentage points across seven conditions, compared to 13~percentage points for the full feature set.

The contributions are:
\begin{itemize}
    \item We separate statistical significance from practical separation using rank-biserial effect size with bootstrap CIs, showing that most ``significant'' features are not deployment-worthy.
    \item Four F1-based descriptors are the only large-effect survivors; they are highly redundant (one latent dimension explains 79\% of variance).
    \item Per-codec-condition evaluation on the official ASVspoof~2021~LA set shows the F1 subset holds nearly constant EER across seven conditions, while the full feature set is unstable under narrowband telephony.
    \item Decision-boundary sensitivity evidence via counterfactual perturbation and speaker-identity robustness via mixed-effects modeling complete the validation.
\end{itemize}

\section{Related Work}
Early spoofing countermeasures relied on hand-crafted cepstral features, but graph-attention architectures like AASIST now achieve EER below 1\% on ASVspoof 2019 LA \cite{jung2022aasist,li2024survey}, and challenge series like ASVspoof have driven much of this progress \cite{todisco2019asvspoof,yamagishi2021asvspoof,liu2023asvspoof}. The 2021 evaluation matters here because it moved beyond clean lab conditions to include compressed and channel-degraded audio, making codec stability an explicit evaluation axis.

\textbf{Neural systems.} Raw-waveform models like RawNet2 \cite{tak2021rawnet2} and self-supervised front ends \cite{guo2024wavlm} improve accuracy over earlier baselines, and style--linguistics mismatch models help generalization \cite{zhu2024slim}. These systems work well but are hard to interpret: when performance drops, it is unclear which acoustic cues broke---the synthesis attack, the codec, or both. Li \etal~\cite{li2024temporal} recently proposed temporal class-activation maps for interpretability in neural anti-spoofing, but the explanation is post-hoc rather than built into the features.

\textbf{Hand-crafted front ends.} PNCCs were designed for robustness to noise and channel variability \cite{kim2016pncc}, while LFCCs remain a solid spoofing baseline \cite{zhou2011lfcc}. Researchers have also improved cross-channel performance by injecting codec variability during training \cite{cohen2022augmentation,chen2021ur}. Yang \etal~\cite{yang2020significance} analyzed which spectral sub-bands carry spoofing cues, finding that high-frequency regions are most discriminative. These studies generally treat the front end as one block inside a larger system, though, and do not ask which low-dimensional descriptors hold up before fusion or augmentation.

\textbf{Formant-based approaches.} Sahidullah \etal~\cite{sahidullah2015comparison} showed that source and vocal-tract features (including formant frequencies and bandwidths) could detect synthetic speech on the ASVspoof 2015 corpus; articulatory information carries spoofing cues. More recently, Cuccovillo \etal~\cite{cuccovillo2023sfat} proposed SFAT-Net, which applies formant analysis via an audio spectrogram transformer, and Negroni \etal~\cite{negroni2026formant} extended this line with a multi-task transformer that models formants for explainable deepfake detection. We screen by effect size rather than reporting all features equally, test codec stability on the official ASVspoof 2021 LA evaluation with per-condition analysis, and show that the surviving signal reduces to a single latent F1 dimension.

\textbf{Physical-signal detection at biometric venues.} Ciftci \etal~\cite{ciftci2020hearts} showed that biological signals (remote photoplethysmography) can detect video deepfakes, a physical-signal framing that parallels our use of vocal-tract resonances for audio deepfakes, trading raw accuracy for interpretability grounded in measurable physiology.

\textbf{Interpretability framing.} Following Rudin \cite{rudin2019stop}, we distinguish inherently interpretable models from post-hoc explanations of black-box systems. Formant features have direct physical meaning tied to vocal-tract resonances \cite{stevens2000acoustic}, satisfying what Lipton \cite{lipton2018mythos} calls ``decomposability'': each input maps to a measurable articulatory property. When paired with a linear classifier (logistic regression achieves comparable EER on these four features; see Section~5), the complete pipeline is inherently interpretable in Rudin's sense. We also evaluate an RBF-SVM as a stronger, though not itself interpretable, performance reference.

\section{Materials, Feature Space, and Analysis Protocol}
\subsection{Data and Deployment Setting}
The clean training data come from the ASVspoof 2019 logical access (LA) corpus \cite{todisco2019asvspoof}. The training partition contains 25,380 utterances: 2,580 bona fide recordings and 22,800 spoofed examples generated by six synthesis/conversion systems (A01--A06; see Table~\ref{tab:eval_conditions}, Panel~A). The class ratio is approximately 1:8.8; all classifiers use balanced class weights and we report equal error rate (EER) throughout.

Deployment robustness is evaluated on the \textbf{official ASVspoof 2021 LA evaluation set} \cite{yamagishi2021asvspoof,liu2023asvspoof}, which contains 181,566 utterances (18,452 bona fide, 163,114 spoofed) generated by 13 unseen attack conditions (A07--A19) and transmitted through seven codec/channel conditions. Table~\ref{tab:eval_conditions} summarizes both dimensions. Each codec/channel condition contains exactly 25,938 utterances. The evaluation differs from training in two ways---unseen spoofing algorithms and channel/codec degradation---and, following the ASVspoof~2021 protocol, we do not retrain or tune on 2021 data.

\begin{table}[t]
\caption{Attack systems and codec/channel conditions. Training attacks (A01--A06) follow \cite{todisco2019asvspoof}; evaluation attacks (A07--A19) follow \cite{liu2023asvspoof}. Exact A07--A19 architectures were not disclosed to challenge participants. Codec standards: G.711 \cite{itu_g711}.}
\label{tab:eval_conditions}
\centering
\footnotesize
\setlength{\tabcolsep}{3pt}
\begin{tabular}{llp{0.42\columnwidth}}
\toprule
\multicolumn{3}{l}{\textbf{Panel A: Attack conditions}} \\
\midrule
ID & Type & Description \\
\midrule
\multicolumn{3}{l}{\emph{Training (A01--A06, seen)}} \\
A01 & TTS & Neural waveform model \\
A02 & TTS & Vocoder-based TTS \\
A03 & TTS & Vocoder-based TTS (different) \\
A04 & TTS & Waveform concatenation \\
A05 & VC & Vocoder-based conversion \\
A06 & VC & Spectral-filtering conversion \\
\midrule
\multicolumn{3}{l}{\emph{Evaluation (A07--A19, unseen)}} \\
A07 & TTS & Neural waveform, vocoder-based \\
A08 & TTS & Neural waveform model \\
A09 & TTS & Traditional vocoder \\
A10 & TTS & Neural waveform model \\
A11 & TTS & Griffin-Lim vocoder \\
A12 & TTS & Neural waveform model \\
A13 & VC & Spectral conversion + filtering \\
A14 & VC & Vocoder-based conversion \\
A15 & VC & Waveform concatenation \\
A16 & TTS & Waveform concatenation \\
A17 & VC & Vocoder-based conversion \\
A18 & TTS & Vocoder-based TTS \\
A19 & TTS & Neural TTS \\
\midrule
\multicolumn{3}{l}{\textbf{Panel B: Codec/channel conditions (evaluation only)}} \\
\midrule
Condition & Band & Standard / notes \\
\midrule
None & 16\,kHz & Uncompressed (attack shift only) \\
G.722 & 16\,kHz & ITU-T wideband codec \\
Opus & 16\,kHz & IETF interactive audio codec \cite{valin2012opus} \\
G.711 A-law & 8\,kHz & ITU-T narrowband PCM \cite{itu_g711} \\
G.711 $\mu$-law & 8\,kHz & ITU-T narrowband PCM \cite{itu_g711} \\
GSM & 8\,kHz & ETSI GSM 06.10 full-rate \\
PSTN & 8\,kHz & Telephone network (end-to-end) \\
\bottomrule
\end{tabular}
\end{table}

As a secondary cross-dataset probe, we evaluate on the full In-the-Wild corpus \cite{muller2022does} (31,779 utterances of publicly available deepfake audio from diverse generators and speakers unrelated to ASVspoof, with matched v4 acoustic feature extraction), testing generalization to fully unseen generators and recording conditions.

Training uses clean audio only; evaluation is intentionally harder, combining unseen attacks with codec degradation. We report per-codec-condition EER to separate wideband (uncompressed, G.722, Opus) from narrowband (G.711, GSM, PSTN) effects.

\subsection{Feature Inventory}
Each utterance is summarized by 36 scalar descriptors divided into four acoustic families (Table~\ref{tab:features} lists all 36 by the exact code names emitted by the released feature-extraction scripts). We use utterance-level summary statistics rather than frame-level sequences because they isolate whether a feature carries a stable real--fake signal that survives codec degradation, and they match the utterance-level decision granularity of deployed systems.

\begin{table}[t]
\caption{The 36 utterance-level descriptors by family, listed by the code names emitted by the released feature-extraction scripts.}
\label{tab:features}
\centering
\footnotesize
\setlength{\tabcolsep}{3pt}
\begin{tabular}{lp{0.04\columnwidth}p{0.67\columnwidth}}
\toprule
Family & $n$ & Descriptors (code names) \\
\midrule
Formant & 8 & \texttt{F1\_mean}, \texttt{F1\_std}, \texttt{F1\_bw}, \texttt{F2\_mean}, \texttt{F2\_std}, \texttt{F2\_bw}, \texttt{F1\_F2\_ratio}, \texttt{formant\_disp} \\
PNCC & 13 & \texttt{PNCC\_0}--\texttt{PNCC\_12} \\
LFCC & 8 & \texttt{LFCC\_0}--\texttt{LFCC\_7} \\
Prosody & 7 & \texttt{F0\_mean}, \texttt{F0\_std}, \texttt{F0\_max}, \texttt{voiced\_ratio}, \texttt{energy\_mean}, \texttt{energy\_std}, \texttt{duration} \\
\midrule
Total & 36 & \\
\bottomrule
\end{tabular}
\end{table}

Formant extraction uses Praat's Burg method via \texttt{parselmouth} (time step: auto, max formants: 5, ceiling: 5500\,Hz, window: 25\,ms, pre-emphasis: 50\,Hz). This yields an effective LPC order of 10. Frames where F1 falls outside 200--1200\,Hz or F2 outside 500--3000\,Hz are excluded, serving as a proxy for voiced-frame selection. PNCCs use a 40-channel gammatone filterbank (80--7500\,Hz) with power-law compression and DCT-II (13 coefficients); LFCCs use 20 linear filters with log-power and DCT-II (8 coefficients); prosodic features use pYIN F0 (75--500\,Hz), RMS energy, voiced-frame ratio, and duration. All 36 descriptors are utterance-level means over 16\,kHz audio. The 5500\,Hz ceiling corresponds to Praat's adult-female default; the adult-male default is 5000\,Hz \cite{boersma2022praat}. Re-extracting F1 at the 5000 and 4500\,Hz ceilings shifts the real--fake effect sizes of F1 mean, bandwidth, and F1/F2 ratio by at most 0.06 (0.055, 0.008, and 0.047, respectively) and leaves the four-feature shortlist unchanged; F1 variability is the most ceiling-sensitive descriptor (up to 0.28). A per-gender detection analysis (Sec.~\ref{sec:speaker}) further shows a negligible male/female EER gap, arguing against a gender-specific tracker artifact. The codec ``robustness'' reported in this study reflects the behavior of the formant tracker on degraded audio, not necessarily the preservation of the physical formant structure itself (see Limitations).

\subsection{Statistical Protocol}
We applied increasingly strict filters, moving from significance to effect size to deployment stability. The sequence is summarized in Table~\ref{tab:roadmap}.

\begin{table*}[t]
\caption{Screening roadmap used to isolate codec-robust features.}
\label{tab:roadmap}
\centering
\small
\setlength{\tabcolsep}{4.8pt}
\begin{tabular}{p{0.10\textwidth}p{0.18\textwidth}p{0.16\textwidth}p{0.20\textwidth}p{0.26\textwidth}}
\toprule
Stage & Question & Statistic / tool & Observed outcome & Decision implication \\
\midrule
Distribution check & Are the feature distributions uniformly well modeled by a Gaussian assumption? & Skew inspection across all 36 descriptors & 33 descriptors are near-normal and 3 are clearly skewed & Use one nonparametric test for all features to avoid assumption mismatch \\
Significance screen & Do bona fide and spoofed utterances differ at all on clean data? & Mann--Whitney $U$ with Benjamini--Hochberg correction & 33 / 36 descriptors are significant after FDR correction & Significance alone is not selective enough \\
Effect-size filter & Which descriptors separate the classes by a practically meaningful margin? & Rank-biserial $r$ with bootstrap 95\% CI (10K resamples) & Only 4 descriptors have large effect ($|r| > 0.5$) & Retain only the four F1 descriptors for deeper analysis \\
Attack consistency & Do the shortlisted features generalize across multiple spoofing systems? & Kruskal--Wallis across A01--A06 & All four shortlisted features differ strongly across the six attacks ($H > 10{,}000$) & The signal is not tied to a single easy training attack \\
Redundancy analysis & Are the shortlisted descriptors complementary or mostly duplicative? & Pearson correlation and PCA (standardized features) & Strong positive correlations; first principal component explains 79\% of the variance & The four ``winners'' largely track one dominant F1-related direction \\
Per-codec probe & Does the signal survive different codec conditions? & Per-condition EER on official ASVspoof 2021 LA eval & F1 subset: 3.4\,pp range. All-36: 13.1\,pp range & F1 features are codec-condition stable \\
Classifier check & Does the screening translate to detection performance? & RBF-SVM on fixed splits; multi-seed ($\times 5$) reproducibility & 4-feat: 17\% $\rightarrow$ 22.9\% (+5.9\,pp). All-36: 9.5\% $\rightarrow$ 23.5\% (+14\,pp) & Feature selection reduces absolute degradation \\
\bottomrule
\end{tabular}
\end{table*}

Of the 36 features, 33 are approximately normal while 3 are clearly skewed. To maintain one consistent nonparametric procedure across all descriptors, we apply the Mann--Whitney $U$ test to every feature when comparing bona fide and spoofed speech. We correct for multiple comparisons using the Benjamini--Hochberg procedure \cite{benjamini1995controlling} at $\alpha=0.05$. Bonferroni correction gives the same result: 33/36 features significant, with only PNCC$_2$, LFCC$_2$, and LFCC$_3$ non-significant (all $|r|<0.01$); most $p$-values are so small ($<10^{-6}$) that the stricter threshold makes no practical difference. We therefore compute the rank-biserial correlation $r$ associated with the Mann--Whitney test and use it as the primary screening criterion for practical separation \cite{wendt1972dealing,kerby2014simple}, applying the conventional large-effect threshold of $|r| \geq 0.5$ \cite{cohen1988statistical}. Intuitively, $r$ is how often a random bona fide utterance exceeds a random spoofed one on the feature. We report bootstrap 95\% CIs on $r$ (10,000 resamples) and verify robustness under class-balanced subsampling (downsampling spoof to match bona fide $n$).

For the shortlisted descriptors, we perform: (i) Kruskal--Wallis tests across the six spoofing systems in the clean training set, (ii) Pearson correlation and principal component analysis (PCA) to measure redundancy, and (iii) matched before/after codec comparisons quantifying how much compression erodes the real--fake separation. For codec robustness, we report the relative gap loss
\begin{equation}
L_c(f)=100\times\frac{|\delta_c(f)|}{|\mu_{\text{real}}(f)-\mu_{\text{spoof}}(f)|},
\end{equation}
where $\delta_c(f)$ is the codec-induced shift for matched clean and coded versions of the same utterance and the denominator is the clean bona fide--spoof separation; lower is better, meaning the codec eats only a small fraction of the discriminative gap.

Finally, we train an SVM (RBF kernel, $C\!=\!1.0$, $\gamma\!=\!\text{scale}$, balanced class weights; scikit-learn defaults selected on the development set) on the fixed challenge split and report EER on development data and the codec-degraded 2021 evaluation; a secondary classifier comparison on the shortlisted set checks whether the conclusion depends on the model family.

\section{Results}
\subsection{Family-Level Screening: Only Formants Survive}
After Benjamini--Hochberg correction, 33 of the 36 descriptors differ significantly between bona fide and spoofed speech, but significance is cheap with 25K samples. Table~\ref{tab:family} shows the family-level outcome: only the formant family produces large-effect survivors ($|r|>0.5$), and exactly half of the formant descriptors survive. None of the 13 PNCC descriptors, 8 LFCC descriptors, or 7 prosodic descriptors reaches the large-effect threshold. Class-balanced subsampling (downsampling spoof to match bona fide $n=2{,}580$) confirms that this result is not driven by class imbalance: the four winners' $|r|$ values change by less than 0.02 under balanced sampling. The same four features survive at any threshold between $|r|=0.31$ and $0.63$---a natural gap between the 4th-ranked feature (F1 variability, $|r|=0.64$) and the 5th (F2 mean, $|r|=0.30$).

\begin{table}[t]
\caption{Family-level survival after effect-size screening.}
\label{tab:family}
\centering
\small
\begin{tabular}{lccc}
\toprule
Family & Total & Large-effect & Survivor rate \\
\midrule
Formants & 8 & 4 & 50\% \\
PNCC & 13 & 0 & 0\% \\
LFCC & 8 & 0 & 0\% \\
Prosody & 7 & 0 & 0\% \\
\bottomrule
\end{tabular}
\end{table}

The likely explanation is margin size: formant survivors start with unusually large clean-data separation, giving them room to absorb deployment shift, while near-boundary features have little tolerance for codec distortion. PNCCs and LFCCs summarize spectral detail that codecs directly manipulate through quantization and bandwidth limitation, and utterance-level prosody is too coarse: modern synthesis matches average pitch and energy while missing fine-grained resonance patterns.

\subsection{The Four Large-Effect Winners}
Only four descriptors pass the large-effect threshold ($|r| > 0.5$), and all four are first-formant measures. Table~\ref{tab:winners} summarizes them.

\begin{table}[t]
\caption{The four large-effect features isolated by statistical screening, ordered by rank-biserial magnitude. Bootstrap 95\% CIs computed from 10,000 resamples.}
\label{tab:winners}
\centering
\footnotesize
\setlength{\tabcolsep}{2pt}
\resizebox{\columnwidth}{!}{%
\begin{tabular}{lccccr}
\toprule
Feature & Real & Fake & Gap & $r$ [95\% CI] & EER \\
\midrule
F1 mean  & $666\!\pm\!71$ & $553\!\pm\!77$ & 113 & .71 [.70,.72] & 19.5\% \\
F1 bw    & $515\!\pm\!87$ & $361\!\pm\!120$ & 154 & .70 [.69,.71] & 21.6\% \\
F1/F2    & $.342\!\pm\!.033$ & $.292\!\pm\!.039$ & .050 & .67 [.65,.68] & 23.0\% \\
F1 var   & $272\!\pm\!25$ & $226\!\pm\!42$ & 46 & .64 [.63,.65] & 17.9\% \\
\bottomrule
\end{tabular}}
\end{table}

All four point the same direction. Compared to spoofed speech, bona fide speech has higher, wider, and more variable F1. In articulatory terms, F1 reflects the lowest vocal-tract resonance, shaped by mouth opening, constriction geometry, and damping \cite{stevens2000acoustic}, and the synthetic signals in this dataset under-represent that physical variability. The strongest codec-resistant cue is thus a resonance pattern tied to speech production, not a generic cepstral artifact.

\begin{figure}[t]
    \centering
    \includegraphics[width=\columnwidth]{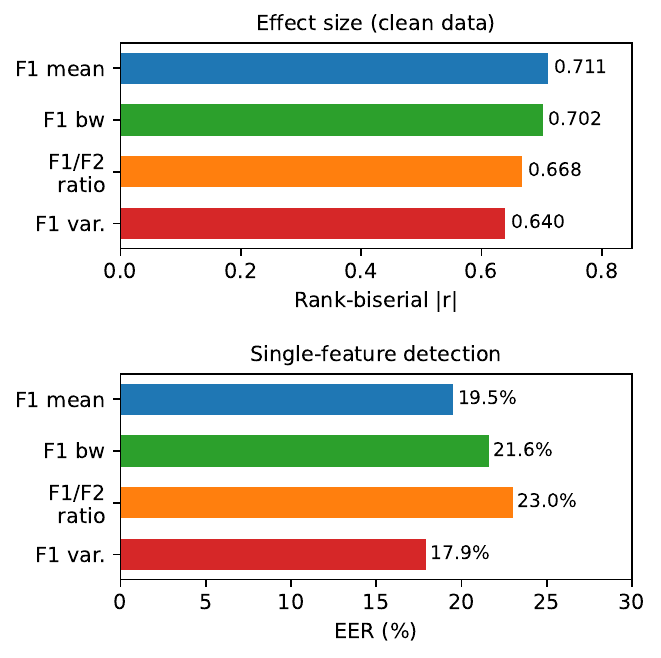}
    \caption{Trade-off view of the four selected descriptors. Top: clean-data rank-biserial correlation. Bottom: EER when each feature is used alone. The same F1 descriptors dominate both views.}
    \label{fig:tradeoffs}
\end{figure}

Figure~\ref{fig:tradeoffs} adds a complementary view: the four top-effect-size descriptors also give the best single-feature EER, with F1 variability leading and the F1/F2 ratio trailing. The ranking is not perfectly monotonic, but the agreement between $r$ and single-feature EER supports effect size as the screening criterion.

\subsection{Attack-Wise Behavior}
For all four F1-based descriptors, the Kruskal--Wallis statistic across attacks A01--A06 exceeds 10,000 ($p\approx 0$), confirming that the signal generalizes across spoofing systems. Per-attack rank-biserial analysis shows that A02 is easiest ($|r|>0.92$ for all four features, F1 variability = 176.7\,Hz vs.\ bona fide 272\,Hz) while A06 is hardest ($|r|<0.13$ for three of four features, F1 mean = 659.6\,Hz vs.\ bona fide 666\,Hz), so F1 behavior tracks the real/fake difference even as attack difficulty varies.

\subsection{The Four Winners Largely Measure One Latent Signal}
\begin{figure}[t]
    \centering
    \includegraphics[width=\columnwidth]{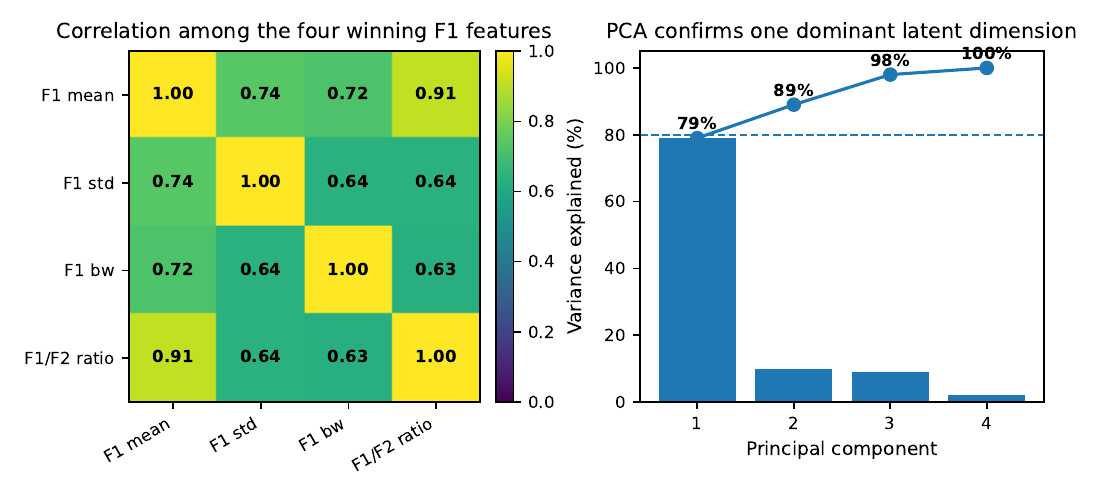}
    \caption{Redundancy analysis for the four large-effect descriptors. Left: Pearson correlations. Right: cumulative PCA explained variance. The first component captures 79\% of the variance, indicating that the four features mostly track the same F1-related direction (within-class PCA confirms: 77\% for spoofed, 56\% for bona fide speech; see text).}
    \label{fig:redundancy}
\end{figure}

The shortlisted descriptors are not independent. Figure~\ref{fig:redundancy} shows their correlation structure and PCA summary on standardized (z-scored) features from the clean training set, both classes pooled. The strongest pair is F1 mean and the F1/F2 ratio ($r=0.91$; Spearman $\rho=0.92$), and all pairwise correlations are positive and moderate to strong. PCA compresses the subset further: PC1 alone explains 79\% of the variance, and the first two components 89\%. PC1 loadings are approximately equal (0.47--0.54), confirming a shared F1-related direction; within-class PCA shows the redundancy is partly between-class (PC1 explains 77\% within spoofed vs.\ 56\% within bona fide speech).

The screening thus keeps landing on one dominant empirical direction---F1 behavior measured from different angles---which lowers the chance of a multiple-testing artifact and suggests the hand-crafted branch could be compressed to one or two latent summaries.

\subsection{Paired Codec Analysis: Streaming Compression Preserves More Than Telephone Coding}
To test whether the F1 signal survives deployment distortions, we run matched before/after codec experiments on the same source utterances encoded with ffmpeg, measuring the codec-induced shift as a percentage of the clean bona fide--spoof separation (a small percentage means the feature remains useful after compression).

\begin{figure}[t]
    \centering
    \includegraphics[width=\columnwidth]{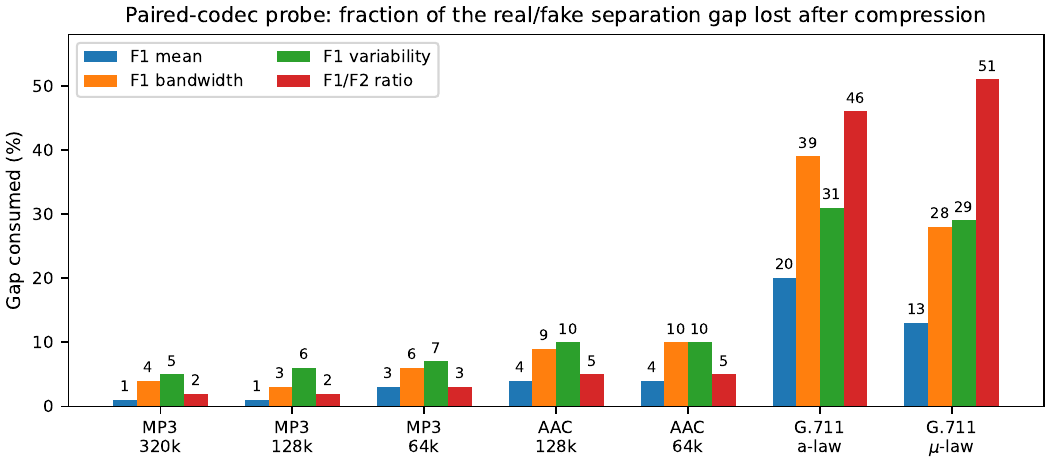}
    \caption{Codec-induced loss measured as a percentage of the clean real--fake separation. Lower is better. MP3/AAC preserve most of the F1 separation, while G.711 erodes a much larger fraction. Note: this paired analysis uses ffmpeg-applied codecs on ASVspoof 2019 LA training material; Table~\ref{tab:percodec} uses the official ASVspoof 2021 LA evaluation conditions.}
    \label{fig:codec}
\end{figure}

Figure~\ref{fig:codec} shows the divide. MP3 and AAC consume only 1--10\% of the original gap for most features, meaning moderate streaming compression largely preserves the discriminative F1 structure. G.711 telephone coding, by contrast, consumes 13--51\% depending on the feature, with the F1/F2 ratio and F1 bandwidth hit hardest---narrowing, but not erasing, the safety margin.

\subsection{Feature Selection Yields Codec-Condition Stability}
The screening translates to detection performance. Table~\ref{tab:percodec} reports per-codec-condition EER for the four-feature and all-feature RBF-SVMs trained on clean 2019 data and evaluated on the official ASVspoof~2021~LA set.

\begin{table}[t]
\caption{Per-codec-condition EER (\%) on ASVspoof 2021 LA eval (unseen attacks A07--A19 + codec conditions). All systems trained on clean 2019 LA data only. The ``none'' row isolates the attack-generalization effect; remaining rows show the combined attack + codec effect.}
\label{tab:percodec}
\centering
\small
\begin{tabular}{lccc}
\toprule
Condition & AASIST & F1 (4) & All 36 \\
\midrule
None (unseen attacks only) & 23.7 & 21.6 & 18.0 \\
\midrule
G.722 (wideband) & 26.0 & 22.4 & 18.2 \\
Opus (wideband) & 36.8 & 21.8 & 20.8 \\
\midrule
G.711 A-law & 23.9 & 23.1 & 25.0 \\
G.711 $\mu$-law & 24.5 & 25.0 & 28.1 \\
GSM & 35.8 & 22.3 & 24.1 \\
PSTN & 46.7 & 21.6 & 31.1 \\
\midrule
\textbf{Overall} & \textbf{30.7} & \textbf{22.9} & \textbf{23.5} \\
Codec-induced range & 23.0\,pp & 3.4\,pp & 13.1\,pp \\
\bottomrule
\end{tabular}
\end{table}

The ``none'' row in Table~\ref{tab:percodec} isolates the attack-generalization effect from codec degradation. AASIST degrades from 1.73\% to 23.7\% (+22\,pp) on unseen attacks alone (full-protocol EER and min~t-DCF for all systems are in Table~\ref{tab:baselines}), whereas the F1-SVM degrades only 1.8\,pp (19.8\%$\to$21.6\%) and all-36 only 1.8\,pp (16.2\%$\to$18.0\%). The small F1 attack-shift partly reflects its higher starting EER, but the attack-shifted EER (21.6\%) stays close to the clean baseline, so F1 features do not exploit attack-specific artifacts that break on unseen systems. On top of this baseline, codecs add further degradation: up to +23\,pp for AASIST (PSTN: 46.7\%), +13\,pp for all-36 (PSTN: 31.1\%), but only +3.4\,pp for the F1 subset ($\mu$-law: 25.0\%), so the F1 features resist both attack-shift and codec-shift.

\begin{figure*}[t]
    \centering
    \begin{minipage}[t]{0.67\textwidth}
        \centering
        \includegraphics[width=\textwidth]{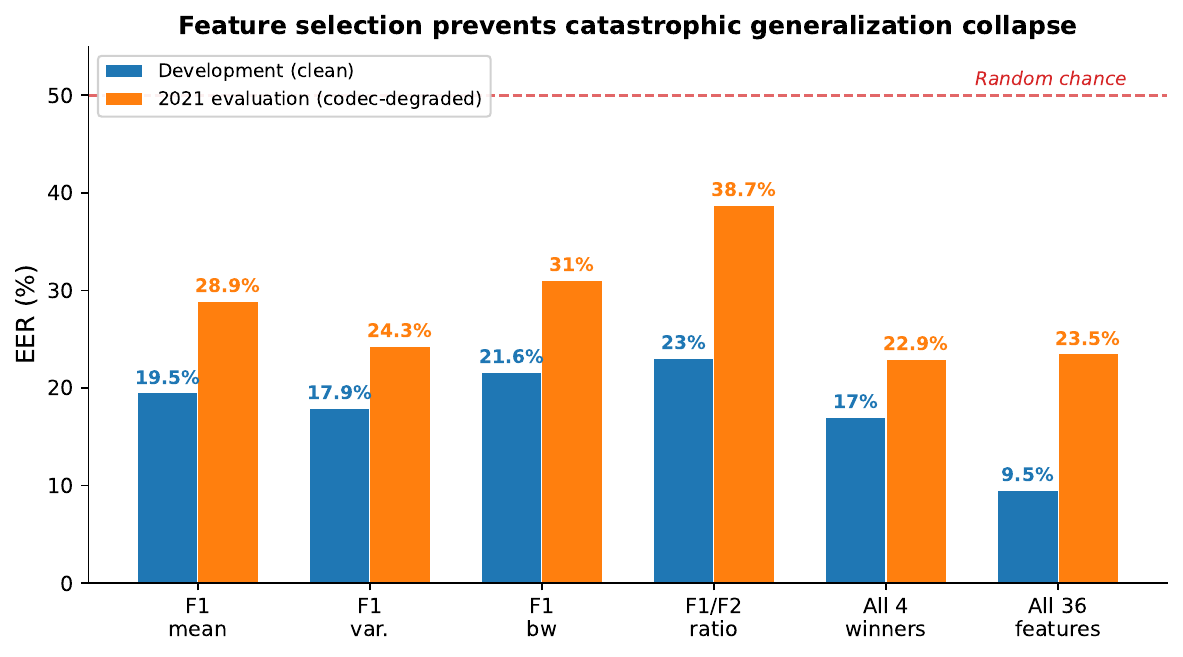}
    \end{minipage}\hfill
    \begin{minipage}[t]{0.29\textwidth}
        \centering
        \includegraphics[width=\textwidth]{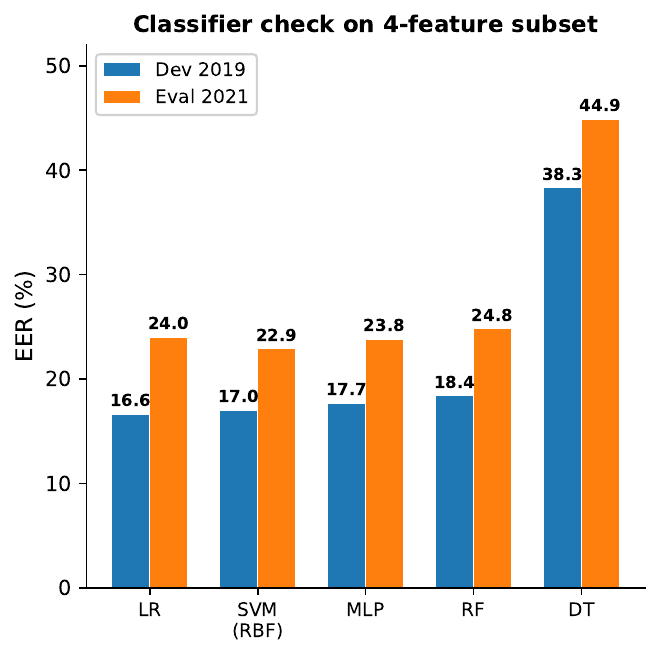}
    \end{minipage}
    \caption{Left: development versus 2021 evaluation EER for different feature subsets. Right: secondary classifier comparison on the four-feature subset; logistic regression, SVM, MLP, and random forest all lie in a narrow band.}
    \label{fig:classifiers}
\end{figure*}

The aggregate EER difference is modest (22.9\% vs.\ 23.5\%); what matters operationally is predictability---the F1 subset holds nearly constant EER across channels while the 36-feature model swings from 18\% (wideband) to 31\% (PSTN). Development-to-evaluation degradation confirms this: the 36-feature model degrades by 14\,pp ($9.5\% \to 23.5\%$), while the F1 subset degrades only 5.9\,pp ($17.0\% \to 22.9\%$), consistent with overfitting to clean-data artifacts.

The right panel of Fig.~\ref{fig:classifiers} shows that the conclusion does not depend on one classifier family. On the four-feature subset, logistic regression, SVM, MLP, and random forest all fall in a narrow 16.6--18.2\% development EER range (multi-seed std $< 0.2\%$), whereas a decision tree degrades sharply.

\subsection{Comparison with Neural Baselines}
Table~\ref{tab:baselines} places the F1-SVM alongside published ASVspoof results. AASIST is the only neural baseline evaluated under the same protocol as our F1-SVM (trained on ASVspoof 2019 LA, evaluated per codec/channel condition on ASVspoof 2021 LA); RawNet2 and LFCC-GMM appear only as published 2019 LA reference points. Table~\ref{tab:baselines} should therefore not be read as a complete robustness ranking across detector families.

\begin{table}[t]
\caption{Full-protocol evaluation. All systems trained on ASVspoof 2019 LA train only. The minimum tandem detection cost function (min~t-DCF) \cite{kinnunen2020tdcf} measures tandem impact on automatic speaker verification. $^\dagger$Published values from original papers. $^\ddagger$Our retrained model (100 epochs, Adam, cosine LR, batch 24); the original paper reports 0.83\%/0.028 on 2019 LA eval.}
\label{tab:baselines}
\centering
\small
\setlength{\tabcolsep}{5pt}
\begin{tabular}{lccccc}
\toprule
System & Dim & \multicolumn{2}{c}{2019 LA eval} & 2021 LA & ITW \\
 & & EER & t-DCF & EER & EER \\
\midrule
AASIST$^\ddagger$ \cite{jung2022aasist} & raw & \textbf{1.73} & 0.056 & 30.7 & 37.6 \\
RawNet2$^\dagger$ \cite{tak2021rawnet2} & raw & 5.13 & 0.118 & --- & --- \\
LFCC-GMM$^\dagger$ \cite{todisco2019asvspoof} & 60 & 8.09 & 0.212 & --- & --- \\
\midrule
All-36 SVM & 36 & 16.2 & 0.337 & 23.5 & 50.0 \\
F1-SVM & 4 & 19.8 & 0.388 & \textbf{22.9} & \textbf{34.7} \\
\bottomrule
\end{tabular}
\end{table}

AASIST was retrained from scratch; its $\sim$30\% 2021 LA EER is typical without augmentation \cite{tak2022odyssey}. AASIST degrades from 1.73\% to 30.7\% ($\sim$18$\times$) on codec-shifted data, while the F1-SVM degrades only from 19.8\% to 22.9\% ($\sim$1.2$\times$). On In-the-Wild \cite{muller2022does} (31,779 files), the F1-SVM achieves 34.7\% EER while the all-36 model collapses to chance (50.0\%).

\section{Discussion}

\subsection{Counterfactual Probe: Decision-Boundary Sensitivity}
To check whether the SVM's decision boundary is geometrically aligned with the F1 direction, we shift each correctly classified bona fide sample's F1 features toward the spoof distribution by the observed class gap (e.g., $-113$\,Hz for F1 mean) in standardized feature space and measure the fraction that flip to ``spoof.'' Shifting all four features at once causes 76.4\% of bona fide samples ($n = 2{,}036$ correctly classified) to flip, compared to only 5.1\% under random-direction perturbation of equal norm (100 draws). Individual flips are: F1 bandwidth 44.3\%, F1/F2 ratio 14.8\%, F1 variability 9.4\%, and F1 mean 7.1\%; these are approximately additive (sum 75.6\% vs.\ combined 76.4\%), so the four features act on largely non-overlapping samples near the boundary. This confirms the SVM boundary is directionally aligned with the F1 direction.

\subsection{Speaker-Level Analysis}
\label{sec:speaker}
ASVspoof 2019 uses only 20 speakers, but spoofed utterances are generated from those same speakers' source recordings, so F1 differences reflect synthesis-induced distortion rather than identity confounding. As a primary test, we compute per-speaker median F1 values for bona fide and spoofed utterances and apply paired Wilcoxon signed-rank tests across the 20 matched pairs. All four F1 features reach significance ($p < 0.001$), confirming that the within-speaker real--fake difference is consistent across the speaker pool. As a secondary sensitivity analysis, a mixed-effects model (speaker as random intercept, class label as fixed effect; fitted via REML in \texttt{statsmodels MixedLM}) shows all four features remain significant ($p < 10^{-10}$, fixed-effect coefficients: F1 mean $\beta = -113$\,Hz, F1 bandwidth $\beta = -154$\,Hz). With only 20 Level-2 clusters, random-effect estimates may be imprecise \cite{maas2005sufficient}, but no model showed a singular fit and the very large fixed effects give margin for small-sample error. Cross-dataset In-the-Wild evaluation (34.7\% EER, unrelated speakers and generators) provides out-of-sample evidence that the signal generalizes beyond the 20 training speakers. Using the official ASVspoof gender labels, the F1-SVM shows almost no gender gap on the 2019 LA evaluation set: 19.7\% EER on male versus 19.8\% on female trials (a 0.1\,pp difference), which argues against a gender-driven tracker artifact. Because these tests use utterance-level summaries, they do not fully control phonetic content; a phoneme-conditioned analysis is left to future work.

\subsection{What the results mean in practice}
\textbf{More features can be worse under deployment shift.} The 36-feature model wins on clean data because extra dimensions let it carve a detailed boundary around known attacks; once codec distortion moves the test data, that boundary no longer aligns with the stable signal. The F1 subset acts as an implicit regularizer, relying on the few cues that stay large-margin after compression, so it need not beat the full model on clean data to be more useful under deployment shift.

\textbf{Stability over absolute performance.} The F1 subset's much narrower codec-condition range (Table~\ref{tab:percodec}) lets it be calibrated once for heterogeneous deployment. Codec-augmented training (injecting G.711 A/$\mu$-law, GSM, and Opus) narrowed the all-36 range to $\sim$2\,pp but did not help the F1 subset (22.9\%$\to$23.1\%); since three of these codecs overlap the evaluation set, this shows robustness to trained codecs rather than unseen-codec generalization, whereas F1 is comparably stable without augmentation.

\textbf{Physical interpretability.} The four descriptors track vocal-tract resonances \cite{stevens2000acoustic} with direct physical meaning. A logistic regression on the same features achieves comparable development EER (16.6\% vs.\ 17.0\%) and 24.0\% on 2021 LA (vs.\ SVM's 22.9\%), so a fully interpretable linear rule can use the F1 signal: a forensic examiner can inspect whether a flagged sample shows unusually narrow, low-variance F1 behavior rather than receiving only an opaque score.

\textbf{Narrowband telephony is the hard case.} A bandwidth-limitation control (16\,kHz$\to$8\,kHz$\to$16\,kHz resampling without codec) shifts F1 mean by $\sim$29\% ($r=0.37$). Narrowband degradation is partly a bandwidth artifact, not purely compression damage. The F1 branch is thus most reliable for wideband or streaming audio and should be treated as a lower-confidence signal on narrowband telephone channels, paired with codec-aware augmentation or channel detection.

\textbf{Positioning relative to neural systems.} F1 features may complement rather than replace a neural detector: SSL-augmented systems \cite{tak2022odyssey} reach far better absolute accuracy ($<$1\% EER), but F1 statistics can tell a human examiner what is acoustically anomalous about a flagged file---and a strong neural ``spoof'' score with normal-looking F1 behavior may itself flag codec or channel mismatch rather than a genuine synthesis artifact. We do not train or evaluate a fused system here; a concrete fusion benefit is left to future work.

\section{Limitations and Future Work}
\textbf{Detection performance.} The F1-SVM operates at 20--25\% EER, far above neural state of the art. It should be used for diagnostic analysis or as a complement to neural systems, not as a standalone detector.

\textbf{Baseline scope.} AASIST is our matched neural anchor under an identical train/eval protocol; RawNet2 and LFCC-GMM are cited only as published 2019 LA reference points. A full per-protocol comparison against detectors explicitly trained for codec/channel robustness is out of scope here and left to future work.

\textbf{Speaker diversity.} With only 20 training speakers, demographic confounding cannot be fully ruled out, though In-the-Wild evaluation (34.7\% EER vs.\ 50.0\% for all-36 and 37.6\% for AASIST) suggests the F1 signal generalizes.

\textbf{Tracker confounds.} The reported codec stability reflects Praat's Burg tracker on degraded audio. Forensic studies report F1 as the least codec-affected formant across two independent trackers \cite{kaiser2018gsm}, which makes a purely tracker-driven explanation less likely; still, controlled resynthesis with known formant targets would be needed to separate genuine formant preservation from tracker estimation error.

\textbf{Generator evolution.} If newer TTS/VC systems model vocal-tract resonances with higher fidelity, the F1 gap may close.

\textbf{Future directions.} Next steps include hybrid fusion with neural detectors, codec-augmented training \cite{cohen2022augmentation}, evaluation on ASVspoof~5 \cite{wang2025asvspoof5}, and broader demographic fairness analysis beyond the binary gender labels used here. Code and pre-computed results are in the supplementary material; datasets are available from the ASVspoof consortium \cite{todisco2019asvspoof,liu2023asvspoof} and Zenodo \cite{muller2022does}.

\section{Conclusion}
Among 36 utterance-level acoustic descriptors evaluated on the official ASVspoof~2021~LA set, only four---all first-formant measures---survive effect-size screening ($|r|>0.5$) and deliver stable detection across seven codec/channel conditions (3.4\,pp EER range vs.\ 13.1\,pp for the full set). A counterfactual probe shows the SVM boundary is aligned with the F1 direction, and paired speaker-level tests with mixed-effects analysis show the F1 signal persists after controlling for speaker identity. These F1 statistics offer a codec-stable diagnostic signal that may complement high-accuracy neural detectors by telling a human examiner what is acoustically different about a flagged file. Which features survive deployment deserves as much attention as pushing EER lower.

\section*{Acknowledgments}
This material is based upon work supported by the Center for Identification Technology and National Science Foundation under Grants 2413228 and 2501916.

{\small
\bibliographystyle{ieee}
\bibliography{refs}
}

\end{document}